\documentclass[]{zakoproc}
\usepackage{graphicx}
\begin{document}

\title{Future Observations of Cosmic Magnetic Fields with the SKA and its Precursors}
\author{Rainer Beck}
\institute{Max-Planck-Institut f\"ur Radioastronomie, Auf dem
H\"ugel 69, 53121 Bonn, Germany} \markboth{R. Beck}{Future
Observations of Cosmic Magnetic Fields \ldots}

\maketitle

\begin{abstract}
The origin of magnetic fields in the Universe is an open problem in
astrophysics and fundamental physics. Polarization observations with
the forthcoming large radio telescopes, especially the Square
Kilometre Array (SKA), will open a new era in the observation of
magnetic fields and should help to understand their origin.
Low-frequency radio synchrotron emission, to be observed with LOFAR,
MWA and the SKA, traces low-energy cosmic ray electrons and allows
us to map the structure of weak magnetic fields in the outer regions
and halos of galaxies, in halos and relics of galaxy clusters and in
the Milky Way. Polarization at higher frequencies (1--10\,GHz), to
be observed with the SKA and its precursors ASKAP and MeerKAT, will
trace magnetic fields in the disks and central regions of galaxies
and in cluster relics  in unprecedented detail. All-sky surveys of
Faraday rotation measures towards a dense grid of polarized
background sources with ASKAP (project POSSUM) and the SKA are
dedicated to measure magnetic fields in intervening galaxies,
clusters and intergalactic filaments, and will be used to model the
overall structure and strength of magnetic fields in the Milky Way.
``Cosmic Magnetism'' is key science for LOFAR, ASKAP and the SKA.

\end{abstract}

\section{Introduction}

The Square Kilometre Array (SKA) is the most ambitious radio
telescope ever planned. With a collecting area of about one square
kilometer, the SKA will be about ten times more sensitive than the
largest single dish telescope (305\,m diameter) at Arecibo (Puerto
Rico), and fifty times more sensitive than the currently most
powerful interferometer, the Expanded Very Large Array (EVLA, at
Socorro/USA). The SKA will continuously cover most of the frequency
range accessible from ground, from 70\,MHz to 10\,GHz in the first
and second phases, later to be extended to at least 25\,GHz. The
third major improvement is the enormously wide field of view,
ranging from 200 square degrees at 70\,MHz to at least 1 square
degree at 1.4\,GHz. The speed to survey a large part of the sky,
particularly at the lower frequencies, will hence be ten thousand to
a million times faster than what is possible today. The SKA is
dedicated to constrain fundamental physics on the dark energy,
gravitation and magnetism.

\section{Technical design of the SKA}

\begin{figure*}[t]
\vspace*{2mm}
\begin{minipage}[t]{6.7cm}
\begin{center}
\includegraphics[bb = 47 47 522 317,width=6.7cm,clip=]{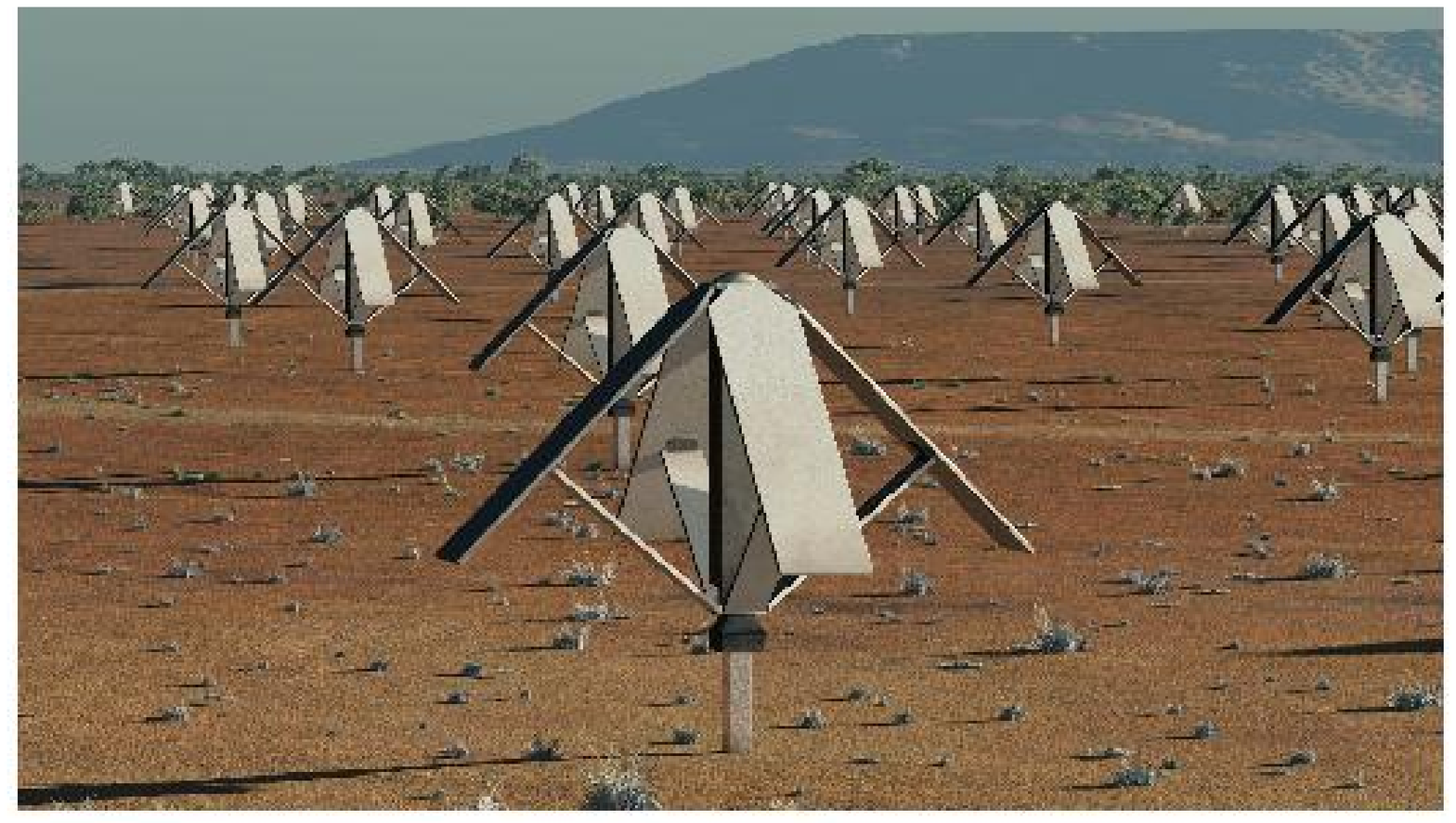}
\caption{SKA sparse aperture array station of dipole elements for
about 70--450\,MHz. Graphics: Swinburne Astronomy Productions and
SKA Project Office (SPO).} \label{fig:low}
\end{center}
\end{minipage}\hfill
\begin{minipage}[t]{6.7cm}
\begin{center}
\includegraphics[bb = 47 47 522 317,width=6.7cm,clip=]{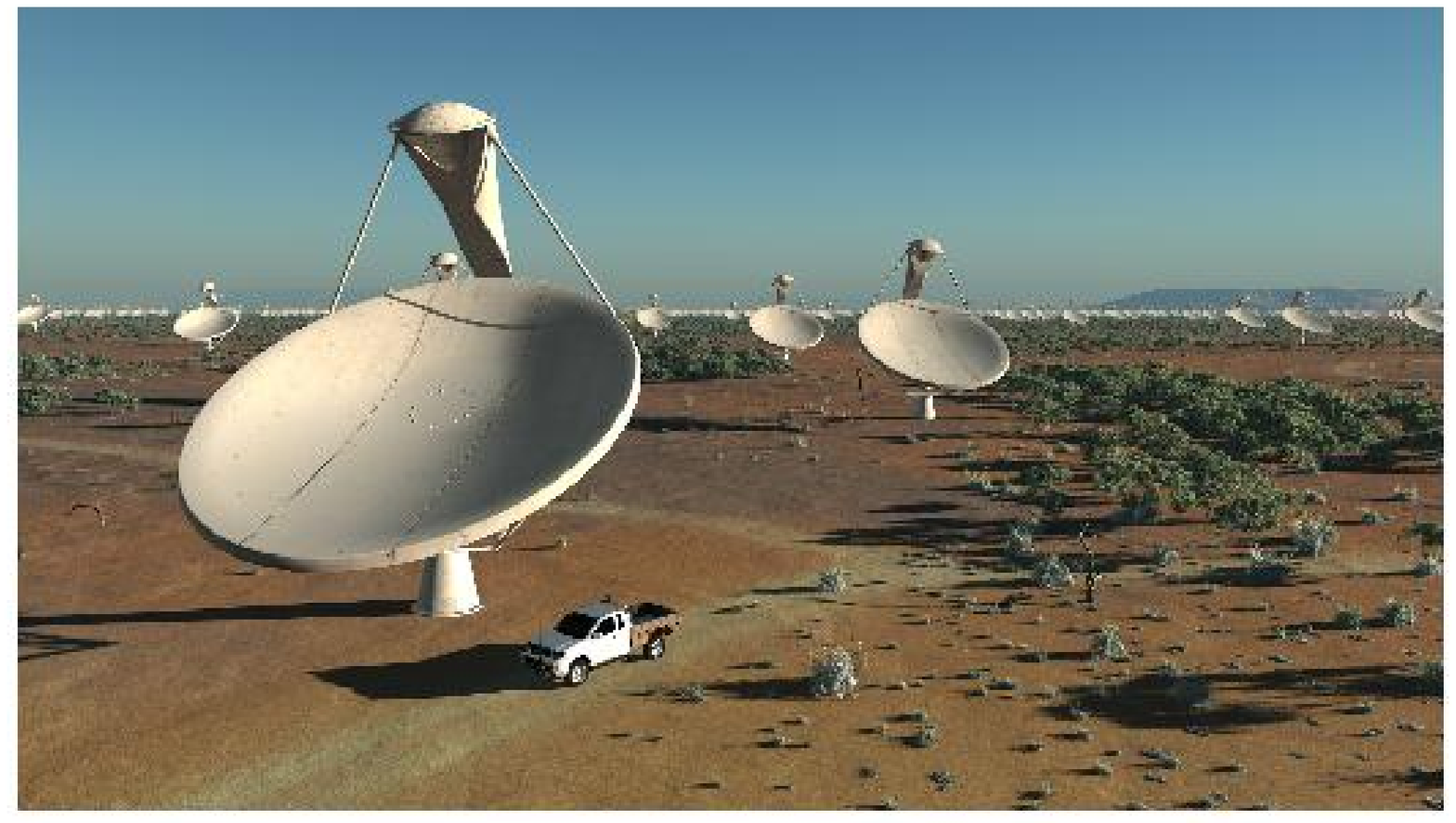}
\caption{SKA parabolic dishes for about 450\,MHz--3\,GHz. Graphics:
Swinburne Astronomy Productions and SPO.} \label{fig:high}
\end{center}
\end{minipage}
\end{figure*}

\begin{figure*}[t]
\vspace*{2mm}
\begin{minipage}[t]{6.7cm}
\begin{center}
\includegraphics[bb = 47 47 522 317,width=6.7cm,clip=]{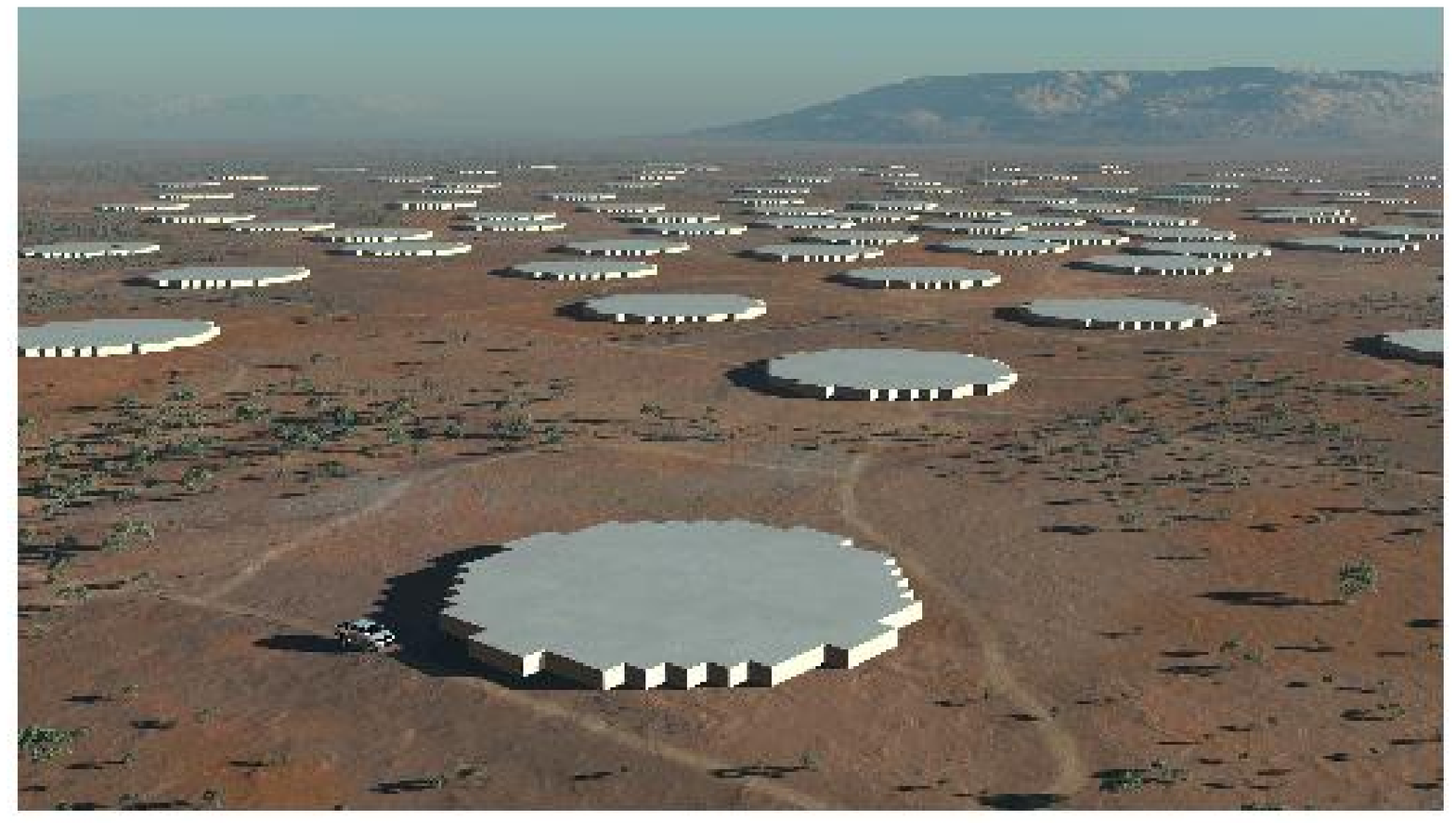}
\caption{SKA dense aperture array station made up of 3\,m x 3\,m
``tiles'' for about 500\,MHz--1\,GHz. Graphics: Swinburne Astronomy
Productions and SPO.} \label{fig:medium}
\end{center}
\end{minipage}\hfill
\begin{minipage}[t]{6.7cm}
\begin{center}
\includegraphics[bb = 47 47 522 317,width=6.7cm,clip=]{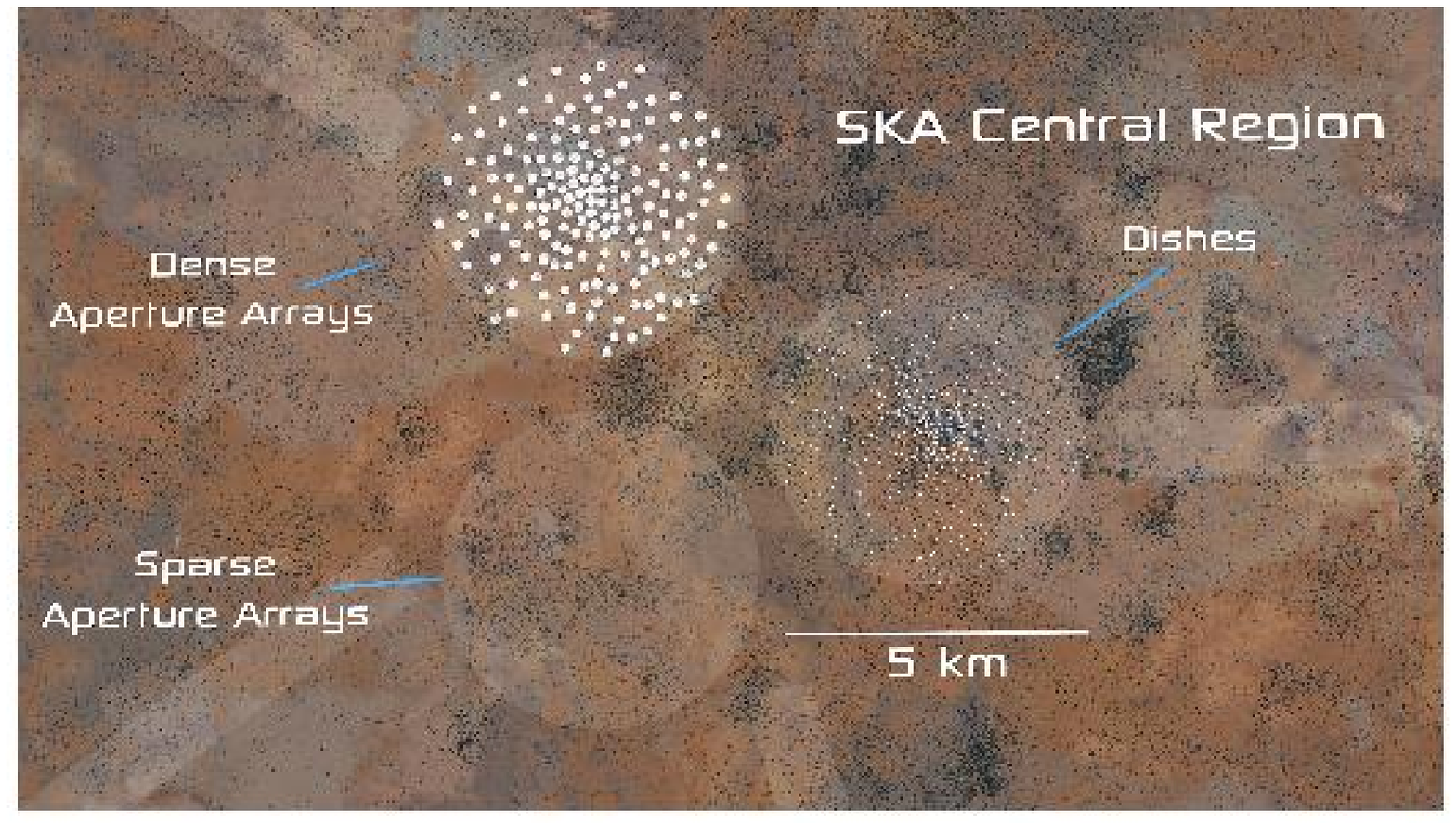}
\caption{SKA core stations for the mid-frequency aperture array and
for the dish array (to be built in South Africa) and the
low-frequency array (to be built in Australia). Graphics: Swinburne
Astronomy Productions and SPO.} \label{fig:core}
\end{center}
\end{minipage}
\end{figure*}

The SKA will be a radio interferometer and consist of many antennas
which are spread over a large area to obtain high the resolving
power. The three separate SLA core regions of 5\,km diameter each
will contain about 50\% of the total collecting area and comprise
dish antennas and the two types of aperture arrays
(Fig.~\ref{fig:core}). The mid-region out to about 180\,km radius
from the core comprises dishes (Fig.~\ref{fig:high}) and sparse
aperture array antennas (Fig.~\ref{fig:low}) aggregated into
stations distributed on a spiral arm pattern. Remote stations with
about 20 dish antennas each will spread out to distances of at least
3000\,km from the core and located on continuations of the spiral
arm pattern. The overall extent of the array determines the angular
resolution, which will be about 0.1\arcsec\ at 100\,MHz and
0.001\arcsec\ at 10\,GHz.

To meet the ambitious specifications and keep the cost to a level
the international community can support, planning and construction
of the SKA requires many technological innovations such as light and
low-cost antennas, detector arrays with a wide field of view,
low-noise amplifiers, high-capacity data transfer, high-speed
parallel-processing computers and high-capacity data storage units.
The enormous data rates of the SKA will demand online image
production with automatic software pipelines.

The frequency range spanning more than two decades cannot be
realized with one single antenna design, so this will be achieved
with a combination of different types of antennas. Under
investigation are the following designs for the low and
mid-frequency ranges:

1. An aperture array of simple dipole antennas with wide spacings (a
``sparse aperture array'') for the low-frequency range (about
70--450\,MHz) (Fig.~\ref{fig:low}). This is a software telescope
with no moving parts, steered solely by electronic phase delays. It
has a large field of view and can observe towards several directions
simultaneously.

2. An array of several thousand parabolic dishes of about 15 meters
diameter each for the medium-frequency range (about
450\,MHz--3\,GHz), each equipped with a wide-bandwidth single-pixel
``feed'' (Fig.~\ref{fig:high}). The surface accuracy of these dishes
will allow a later receiver upgrade to higher frequencies.

As an ``Advanced Instrumentation Programme'' for the full SKA, two
additional technologies for substantially enhancing the field of
view in the 500--1000\,MHz range are under development: aperture
arrays with dense spacings, forming an almost circular station 60\,m
across (Fig.~\ref{fig:medium}) and phased-array feeds for the
parabolic dishes (see below).

\section{Technical developments}

Technical developments around the world are being coordinated by the
SKA Science and Engineering Committee and its executive arm, the SKA
Project Office. The technical work itself is funded from national
and regional sources, and is being carried out via a series of
verification programs. The global coordination is supported by funds
from the European Commission under a program called PrepSKA, the
Preparatory Phase for SKA, whose primary goals are to provide a
costed system design and an implementation plan for the telescope by
2012.

A number of telescopes provide examples of low frequency arrays,
such as the European LOFAR (Low Frequency Array) telescope, with its
core in the Netherlands, the MWA (Murchison Widefield Array) in
Australia, PAPER (Precision Array to Probe the Epoch of
Reionization), also in Australia, and the LWA (Long Wavelength
Array) in the USA. All these long wavelength telescopes are software
telescopes steered by electronic phase delays (phased aperture
array). Examples of dishes with a single-pixel feed are under
development in South Africa (MeerKAT, Karoo Array Telescope).

Dense aperture arrays comprise up to millions of receiving elements
in planar arrays on the ground (Fig.~\ref{fig:medium}) which can be
phased together to point in any direction on the sky. Due to the
large reception pattern of the basic elements, the field of view can
be up to 250 square degrees. This technology can also be adapted to
the focal plane of parabolic dishes. Prototypes of such wide-field
cameras are under construction in Australia (ASKAP, Australian SKA
Pathfinder), the Netherlands (APERTIF) and in Canada (PHAD).

The data from all stations have to be transmitted to a central
computer and processed online. Compared to LOFAR with a data rate of
about 150\,Gigabits per second and a central processing power of
27\,Tflops, the SKA will produce much more data and need much more
processing power - by a factor of at least one hundred. Following
``Moore's law'' of increasing computing power, a processor with
sufficient power should be available by the next decade. The energy
consumption for the computers and cooling will be tens of MegaWatts.

\section{SKA timeline}

The detailed design for low and mid frequencies will be ready until
2013. The development of technologies for the high-frequency band
will start in 2013. Construction of the SKA is planned to start in
2016. In the first phase (until about 2020) about 10\% of the SKA
will be erected (SKA$_1$) (Garrett et al. \cite{garrett10}), with
completion of construction at the low and mid frequency bands
(SKA$_2$) by about 2024, followed by construction at the high band.

The members of the SKA Organisation agreed on a dual site solution
for the SKA with two candidate sites fulfilling the scientific and
logistical requirements: Southern Africa, extending from South
Africa, with a core in the Karoo desert, eastward to Madagascar and
Mauritius and northward into the continent, and Australia, with the
core in Western Australia. The dishes of SKA$_1$ will be built in
South Africa, combined with the MeerKAT telescope, and further
dishes will be added to the ASKAP array in Australia. All the dishes
and the mid-frequency dense aperture array for SKA$_2$ will be built
in Southern Africa. The low-frequency sparse aperture array of
dipole antennas for SKA$_1$ and SKA$_2$ will be built in Australia.


\section{Key science projects}

Apart from the expected technological spin-offs, five main science
questions (Key Science Projects) drive the SKA (Carilli \& Rawlings
\cite{carilli04}).

\begin{figure*}[t]
\vspace*{2mm}
\begin{minipage}[t]{6.2cm}
\begin{center}
\includegraphics[bb = 34 35 265 258,width=6.2cm,clip=]{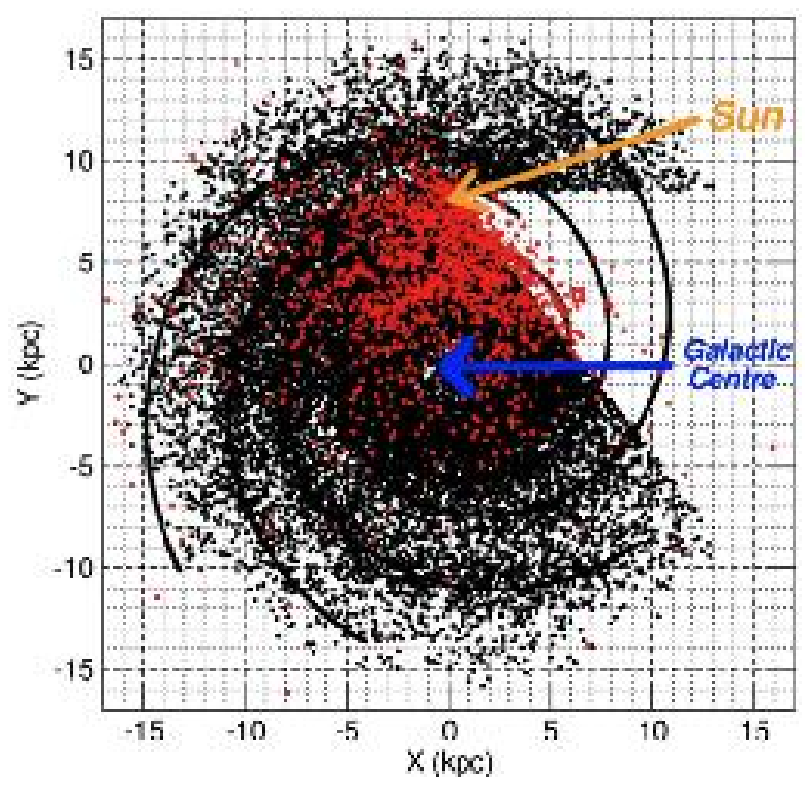}
\caption{Known pulsars in the Milky Way (red) and pulsars expected
with the SKA (black). Simulation: Michael Kramer, MPIfR Bonn.}
\label{fig:pulsars}
\end{center}
\end{minipage}\hfill
\begin{minipage}[t]{7.2cm}
\begin{center}
\includegraphics[width=7.2cm]{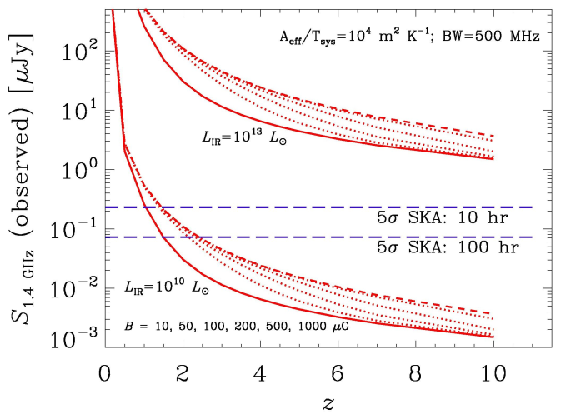}
\caption{Total synchrotron emission of galaxies at 1.4\,GHz as a
function of redshift $z$ and magnetic field strength $B$, and the
$5\sigma$ detection limits for 10\,h and 100\,h integration time
with the SKA (Murphy \cite{murphy09}).} \label{fig:sens}
\end{center}
\end{minipage}
\end{figure*}

\begin{itemize}
\item {Probing the dark ages}

The SKA will use the emission of neutral hydrogen to observe the
most distant objects in the Universe. The energy output from the
first energetic stars and the jets launched near young black holes
(quasars) started to heat the neutral gas, forming bubbles of
ionized gas as structure emerged. This is called the Epoch or
Reionization. The signatures from this exciting transition phase
should still be observable with help of the HI radio line,
redshifted by a factor of about 10. The lowest SKA frequency will
allow us to detect hydrogen at redshifts of up to 20, to search for
the transition from a neutral to an ionized Universe, and hence
provide a critical test of our present-day cosmological model.

\item{Galaxy evolution, cosmology, and dark energy}

The expansion of the Universe is currently accelerating, a not
understood phenomenon, named ``dark energy''. One important method
of distinguishing between the various explanations is to compare the
distribution of galaxies at different epochs in the evolution of the
Universe to the distribution of matter at the time when the Cosmic
Microwave Background (CMB) was formed. Small distortions in the
distribution of matter, called baryon acoustic oscillations, should
persist from the era of CMB formation until today. Tracking if and
how these ripples change in size and spacing over cosmic time can
then tell us if one of the existing models for dark energy is
correct or if a new idea is needed. A deep all-sky SKA survey will
detect hydrogen emission from Milky Way-like galaxies out to
redshifts of about 1. The galaxy observations will be ``sliced'' in
different redshift (time) intervals and hence reveal a comprehensive
picture of the Universe's history.

The same data set will give us unique information about the
evolution of galaxies, how the hydrogen gas was concentrated to form
galaxies, how fast it was transformed into stars, and how much gas
did galaxies acquire during their lifetime from intergalactic space.
The HI survey will simultaneously give us the synchrotron radiation
intensity of the galaxies which is a measure of their star-formation
rate and magnetic field strength.

\item{Tests of General Relativity and detection of gravitational
waves}

Pulsars are ideal probes for experiments in the strong gravitational
field around black holes have yet been made. We expect that almost
all pulsars in the Milky Way will be detected with the SKA
(Fig.~\ref{fig:pulsars}) plus several 100 bright pulsars in nearby
galaxies. The SKA will search for a radio pulsar orbiting around a
black hole, measure time delays in extremely curved space with much
higher precision than with laboratory experiments and hence probe
the limits of General Relativity.

Regular high-precision observations with the SKA of a network of
pulsars with periods of milliseconds opens the way to detect
gravitational waves with wavelengths of many parsecs, as expected
for example from two massive black holes orbiting each other with a
period of a few years resulting from galaxy mergers in the early
Universe. When such a gravitational wave passes by the Earth, the
nearby space-time changes slightly at a frequency of a few nHz
(about 1 oscillation per 30 years). The wave can be detected as
apparent systematic delays and advances of the pulsar clocks in
particular directions relative to the wave propagation on the sky.

\item{The cradle of life}

The SKA will be able to detect the thermal radio emission from
centimeter-sized ``pebbles'' in protoplanetary systems which are
thought to be the first step in assembling Earth-like planets.
Biomolecules are observable in the radio range. Prebiotic chemistry
- the formation of the molecular building blocks necessary for the
creation of life - occurs in interstellar clouds long before that
cloud collapses to form a new solar system. Finally, the SETI
(Search for Extra Terrestrial Intelligence) project will use the SKA
to find hints of technological activities. Ionospheric radar
experiments similar to those on Earth will be detectable out to
several kpc, and Arecibo-type radar beams, like those that we use to
map our neighbor planets in the solar system, out to as far as a few
10\,kpc.

\item{Origin and evolution of cosmic magnetism}

Synchrotron radiation and Faraday rotation revealed magnetic fields
in our Milky Way, nearby spiral galaxies, and in galaxy clusters,
but little is known about magnetic fields in the intergalactic
medium. Furthermore, the origin and evolution of magnetic fields is
still unknown. The SKA will measure the Faraday rotation towards
several tens of million polarized background sources (mostly
quasars), allowing us to derive the magnetic field structures and
strengths of the intervening objects, such as, the Milky Way,
distant spiral galaxies, clusters of galaxies, and in intergalactic
space -- see below.
\end{itemize}

From the five Key Science Projects two major science goals have been
identified that drive the technical specifications for the first
phase (SKA$_1$):
\begin{itemize}
\item Origins: Understanding the history and role of neutral
hydrogen in the Universe from the dark ages to the present-day.
\item Fundamental Physics: Detecting and timing binary pulsars and
spin-stable millisecond pulsars in order to test theories of
gravity.
\end{itemize}

\section{Future magnetic field observations}

Next-generation radio telescopes will widen the range of observable
magnetic phenomena. At low frequencies, synchrotron emission will be
observed from aging electrons far away from their places of origin.
Low frequencies are also ideal to search for small Faraday rotation
measures from weak interstellar and intergalactic fields (Arshakian
\& Beck \cite{arshakian11}) and in steep-spectrum cluster relics
(Brunetti et al. \cite{brunetti08}). The recently completed LOFAR
(operating at 10--240\,MHz), followed by the MWA and the LWA (both
under construction), are suitable instruments to search for weak
magnetic fields in outer galaxy disks, galaxy halos and cluster
halos. First LOFAR results have been presented at this conference
(Anderson et al., Mulcahy et al., this volume).

LOFAR will detect all pulsars within 2\,kpc of the Sun and discover
about 1000 new nearby pulsars, especially at high latitudes (van
Leeuwen \& Stappers \cite{leeuwen10}). Most of these are expected to
emit strong, linearly polarized signals at low frequencies. This
will allows us to measure their RMs and to derive the magnetic field
structure near to the Sun.

Deep high-resolution observations at high frequencies, where Faraday
effects are small, require a major increase in sensitivity of
continuum observations, to be achieved by the EVLA and the SKA. The
detailed structure of the magnetic fields in the ISM of galaxies, in
galaxy halos, cluster halos and cluster relics will be observed. The
magnetic power spectra can be measured (Vogt \& En{\ss}lin
\cite{vogt05}). Direct insight into the interaction between gas and
magnetic fields in these objects will become possible. The SKA will
also allow to measure the Zeeman effect of weak magnetic fields in
the Milky Way and in nearby galaxies.

Detection of polarized emission from distant, unresolved galaxies
will reveal large-scale ordered fields (Stil et al. \cite{stil09}),
to be compared with the predictions of dynamo theory (Arshakian et
al. \cite{arshakian09}). The SKA will detect Milky-Way type galaxies
at $z\le1.5$ (Murphy \cite{murphy09}) and their polarized emission
at $z\le0.5$ (assuming 10\% polarization). Cluster ``relics'' are
highly polarized (van Weeren et al. \cite{weeren10}) and will also
be detectable at large redshifts.

Bright starburst galaxies are not expected to host ordered fields.
Unpolarized synchrotron emission from starburst galaxies, signature
of turbulent magnetic fields, will be detected with the SKA out to
large redshifts, depending on luminosity and magnetic field strength
(Fig.~\ref{fig:sens}), and from cluster halos. However, for fields
weaker than 3.25\,$\mu$G $(1+z)^2$, energy loss of cosmic-ray
electrons is dominated by the inverse Compton effect with CMB
photons, so that the energy appears mostly in X-rays, not in the
radio range. On the other hand, for strong fields the energy range
of the electrons emitting at a 1.4\,GHz drops to low energies, where
ionization and bremsstrahlung losses become dominant (Murphy
\cite{murphy09}). In summary, the mere detection of synchrotron
emission of galaxies at high redshifts will constrain the range of
allowed magnetic field strengths.

If polarized emission from galaxies, cluster halos or cluster relics
is too weak to be detected, the method of {\em RM grids}\ towards
background QSOs can still be applied and allows us to determine the
field strength and pattern in an intervening galaxy. This method can
be applied to distances of young QSOs ($z\simeq5$). Regular fields
of several $\mu$G strength were already detected in distant galaxies
(Bernet et al. \cite{bernet08}, Kronberg et al. \cite{kronberg08}).
Mean-field dynamo theory predicts RMs from evolving regular fields
with increasing coherence scale at $z\le3$ (Arshakian et al.
\cite{arshakian09}). (Note that the observed RM values are reduced
by the redshift dilution factor of $(1+z)^{-2}$.) A reliable model
for the field structure of nearby galaxies, cluster halos and
cluster relics needs RM values from a large number of polarized
background sources, hence large sensitivity and high survey speed
(Krause et al. \cite{krause09}).

The {\em POSSUM}\ all-sky survey at 1.1--1.4\,GHz with the ASKAP
telescope (under construction) with about 30\,deg$^2$ field of view
will measure about 100\,RMs of extragalactic sources per square
degree within 10~h integration time.

The {\em SKA Magnetism Key Science Project}\ plans to observe a
wide-field survey (at least $10^4$\,deg$^2$) around 1\,GHz with 1\,h
integration per field which will detect sources of 0.5--1\,$\mu$Jy
flux density and measure at least 1500\,RMs\,deg$^{-2}$. This will
contain at least 1.5\,$10^7$ RMs from compact polarized
extragalactic sources at a mean spacing of $\simeq90''$ (Gaensler et
al. \cite{gaensler04}). This survey will be used to model the
structure and strength of the magnetic fields in the Milky Way, in
intervening galaxies and clusters and in the intergalactic medium
(Beck \& Gaensler \cite{beck04}). The SKA pulsar survey will find
about 20,000 new pulsars which will mostly be polarized and reveal
RMs (Fig.~\ref{fig:pulsars}), suited to map the Milky Way's magnetic
field with high precision. More than 10,000 RM values are expected
in the area of the galaxy M~31 and will allow the detailed
reconstruction of the 3-D field structure. Simple patterns of
regular fields can be recognized out to distances of about 100\,Mpc
(Stepanov et al. \cite{stepanov08}) where the polarized emission is
far too low to be mapped. The evolution of field strength in cluster
halos can be measured by the RM grid method to redshifts of about 1
(Krause et al. \cite{krause09}).

If the filaments of the local Cosmic Web outside clusters contain a
magnetic field (Ryu et al. \cite{ryu08}), possibly enhanced by IGM
shocks, we hope to detect this field by direct observation of its
total synchrotron emission (Keshet et al. \cite{keshet04}) and
possibly its polarization, or by Faraday rotation towards background
sources. For fields of $\approx 10^{-8}-10^{-7}$\,G with 1\,Mpc
coherence length and $n_e\approx 10^{-5}$\,cm$^{-3}$ electron
density, Faraday rotation measures between 0.1 and 1\,rad\,m$^{-2}$
are expected which will be challenging to detect even with LOFAR.
More promising is a statistical analysis like the measurement of the
power spectrum of the magnetic field of the Cosmic Web (Kolatt
\cite{kolatt98}) or the cross-correlation with other large-scale
structure indicators like the galaxy density field (Stasyszyn et al.
\cite{stasyszyn10}).

If an overall IGM field with a coherence length of a few Mpc existed
in the early Universe and its strength varied proportional to
$(1+z)^2$, its signature may become evident at redshifts of $z>3$.
Averaging over a large number of RMs is required to unravel the IGM
signal. The goal is to detect an IGM magnetic field of 0.1\,nG,
which needs an RM density of $\approx1000$ sources deg$^{-2}$
(Kolatt \cite{kolatt98}), achievable with the SKA. Detection of a
general IGM field, or placing stringent upper limits on it, will
provide powerful observational constraints on the origin of cosmic
magnetism.


\end{document}